\documentclass[preprint,epsf,aps]{revtex4}
\usepackage{graphicx}
\begin{document}

\title{Resistivity saturation in PrFeAsO$_{1-x}$F$_y$ superconductor: An evidence of strong electron-phonon coupling}

\author{D. Bhoi and P. Mandal}
\affiliation{Saha Institute of Nuclear Physics, 1/AF Bidhannagar,
Calcutta 700 064, India\\}
\author{ P. Choudhury}
\affiliation{Central Glass and Ceramic Research Institute, 196 Raja
S. C. Mullick Road, Calcutta 700 032, India\\}
\date{\today}
\begin{abstract}

We have measured the resistivity of PrFeAsO$_{1-x}$F$_y$ samples
over a wide range of temperature in order to elucidate the role of
electron-phonon interaction  on normal- and superconducting-state
properties. The linear $T$ dependence of $\rho$ above 170 K followed
by a saturationlike behavior at higher temperature is a clear
signature of strong electron-phonon coupling. From the analysis of
$T$ dependence of $\rho$, we have estimated several normal-state
parameters useful for understanding the origin of superconductivity
in this system. Our results suggest that Fe-based oxypnictides are
phonon mediated BCS superconductors like Chevrel phases and A15
compounds.

\vskip 1cm
\end{abstract}
\maketitle
\newpage

{\bf I. INTRODUCTION}

The discovery of iron-based layered oxypnictide LaFeAsO$_{1-x}$F$_x$
with high superconducting transition temperature ($T_c$) 26 K has
stimulated  intense experimental and theoretical activities in the
field of superconductivity \cite{kami1,kami2}. Following the initial
report, attempts have been made to enhance $T_c$ mainly by changing
the sample composition. It has been shown that $T_c$ increases
significantly when La is replaced by other rare earth elements ($R$)
of smaller ionic size such as Ce, Pr, Nd, Sm, etc
\cite{chen1,chen2,ren1,ren2,ren3}. $T_c$ as high as 55 K has been
reached in this class of materials, which is the highest after
cuprate superconductors. The stoichiometric compound $R$FeAsO is a
nonsuperconducting metal, undergoes a structural phase transition
around $T_s$=150 K and exhibits a long-range antiferromagnetic
ordering of Fe moments slightly below $T_s$\cite{kami1,dong}.
Partial substitution of fluorine for oxygen or creating oxygen
vacancy suppresses both structural and magnetic phase transitions,
and drives the system to a superconducting ground state. Besides
high transition temperature, this system exhibits many interesting
phenomena possibly due to its layered structure as in the case of
high-$T_c$ cuprates and the presence of iron. At room-temperature,
the crystal structure of $R$FeAsO is tetragonal with space group
$P4/nmm$ and it consists of $R$O and FeAs layers which are stacked
along the $c$ axis. In spite of high transition temperature and
layered structure, there are some important differences with respect
to cuprate superconductors. For oxypnictides, the parent compound is
not a Mott insulator and the superconducting-state properties are
insensitive to oxygen or fluorine content over a wide range. Another
important difference between high-$T_c$ cuprates and oxypnictides is
the dependence of $T_c$ on the ionic size of rare earth ion. Unlike
high-$T_c$ cuprates, the present system shows that $T_c$ increases
with the decrease of ionic radius of rare earth ion \cite{ren3}.
Moreover, the symmetry  and temperature dependence of gap in these
two systems are not same. The gap function of the latter is $s$-wave
and is consistent with the BCS theory \cite{chen,kondo}. \\

Several theoretical and experimental attempts aimed at identifying
the possible superconducting mechanism in oxypnictides have been
made \cite{chen,kondo,drec,sing,lebe,haul,boer,mazi,esch}. The high
transition temperature, structural similarities with cuprates and
the occurrence of superconductivity on the verge of a ferromagnetic
instability led to speculate that the superconductivity in this
system is unconventional. Most of the theoretical reports suggest
that superconductivity in oxypnictides is not mediated by phonon
\cite{sing,lebe,haul,boer,mazi}. Boeri {\it et al}. \cite{boer}
calculated the value of electron-phonon (e-ph) coupling constant
$\lambda_{tr}$=0.21 for pure LaFeAsO using the standard
Migdal-Eliashberg theory and concluded that $\lambda_{tr}$ should be
even smaller for the fluorine-doped superconducting samples.
According to them, the maximum value of $T_c$  due to the e-ph
coupling should not be above 0.8 K and a 5-6 times larger coupling
constant would be needed to reproduce the experimental value of
$T_c$ in this system. In contrast to this, Eschrig \cite{esch}
argued in favor of strong e-ph coupling of the Fe inplane breathing
mode for the high transition temperature in this system. Chen {\it
et al}. \cite{chen} from the Andreev reflection spectroscopy and
Kondo {\it et al}. \cite{kondo} from the angle resolved
photoemission spectroscopy, have shown s-wave symmetry of the order
parameter as in the case of phonon mediated conventional
superconductors. The temperature dependence of energy gap is also
consistent with the BCS theory. To shed some light on the mechanism
of superconductivity, we have measured the temperature dependence of
resistivity  up to 550 K for PrFeAsO$_{1-x}$F$_y$ samples. From the
analysis of high-temperature resistivity, the e-ph coupling strength
for the superconducting and nonsuperconducting samples has been
estimated. Indeed, we observe that $\lambda_{tr}$ is quite small
(0.24)  for the nonsuperconducting sample in accordance with the
theoretical prediction but it is 5-6 times larger
for the superconducting samples.  \\

{\bf II. SAMPLE PREPARATION AND EXPERIMENTAL TECHNIQUES}

Good quality single phase samples are necessary for understanding
the normal-state transport properties and the mechanism of
superconductivity in this system. Usually, two methods have been
followed for the preparation of iron-based oxypnictides. Using
high-pressure synthesis technique, one can obtain fluorine-free and
oxygen-deficient $R$FeAsO$_{1-\delta}$ superconducting samples
\cite{ren4,hiro}. However, samples prepared in this method contain
appreciable amount of impurity phases resulting from the unreacted
ingredients. In the other method, oxygen is partially replaced by
fluorine and the sample is prepared either in high vacuum or in the
presence of inert gas. Though samples prepared in this method are
superior in quality, they are not always free from impurity phases
\cite{sefa,ding}. We use a slightly different method for the
preparation of superconducting PrFeAsO$_{1-x}$F$_y$ samples. The
nominal composition for these samples is oxygen deficient and at the
same time oxygen is partially replaced by a small amount of fluorine
($x \not=y \not= $0). We observe that single phase samples with
$x$$\leq$0.4 can be prepared in this method. Polycrystalline samples
of nominal compositions PrFeAsO, PrFeAsO$_{0.6}$F$_{0.12}$ and
PrFeAsO$_{0.7}$F$_{0.12}$ were prepared by conventional solid-state
reaction method. High purity chemicals Pr (99.9$\%$), Fe
(99.998$\%$), Fe$_2$O$_3$ (99.99$\%$), As (99.999$\%$), PrF$_3$
(99.9$\%$) and Pr$_6$O$_{11}$ (99.99$\%$) from Alfa-Aesar were used
for the sample preparation. Finely grounded powders of
Pr$_{0.96}$As, Fe, Fe$_2$O$_3$, Pr$_6$O$_{11}$ (pre-heated at 600
$^o$C) and PrF$_3$ were thoroughly mixed in appropriate ratios and
then pressed into pellets. The pellets were wrapped with Ta foil and
sealed in an evacuated quartz tube. They were then annealed at 1250
$^o$C for 36 h. Pr$_{0.96}$As was obtained by slowly reacting Pr
chips and As pieces first at 850 $^o$C for 24 h and then at 950
$^o$C for another 24 h in an evacuated quartz tube. The product was
reground, pressed into pellets and then sealed again in a quartz
tube and heated at 1150 $^o$C for about 24 h. The phase purity and
the room-temperature structural parameters were determined by powder
x-ray diffraction (XRD) method with Cu $K_\alpha$ radiation. dc
magnetization and electrical resistivity measurements up to 300 K
were done using a Quantum Design Physical Property Measurement
System (PPMS). High temperature resistivity above 300 K was measured
in a home-made set-up. Resistivity was measured by standard
four-probe technique. In order to avoid  oxidation,
resistivity measurements at high temperatures were done in vacuum. \\

{\bf III. EXPERIMENTAL RESULTS AND DISCUSSION}

{\bf A. Powder x-ray diffraction analysis}\\

Figure 1 shows the XRD pattern for the fluorine-doped
PrFeAsO$_{0.6}$F$_{0.12}$ sample at room temperature. No impurity
phase was observed for $x$=0.3 and 0.4 samples. The diffraction
pattern can be well indexed on the basis of tetragonal ZrCuSiAs-type
structure with the space group P4/$nmm$.  The lattice parameters
obtained from the Rietveld refinements are $a$=3.9711 $\AA$ and
$c$=8.5815 $\AA$. These values of lattice parameters are comparable
with those reported for fluorine-doped and oxygen-deficient PrFeAsO
samples \cite{ren1,ren4}. As expected, the lattice parameters of the
present samples are located between CeFeAsO$_{1-x}$F$_x$ and
NdFeAsO$_{1-x}$F$_x$ \cite{chen2,hiro}. The deduced bond lengths are
$d$(Pr-O/F)=2.3405, $d$(Pr-As)=3.3008, $d$(As-Fe)=2.3828, and
$d$(Fe-Fe)=2.8081 $\AA$. The values of several bond angles are
$\langle${O/F-Pr-As}$\rangle$=76.57,
$\langle${O/F-Pr-O/F}$\rangle$=73.73,
$\langle${Pr-As-Fe}$\rangle$=77.83,
$\langle${Fe-As-Fe}$\rangle$=72.20, and
$\langle${As-Fe-As}$\rangle$=107.79$^{o}$. We would like to mention
here that the Fe-Fe distance in the present system is about 1.4 $\%$
smaller than the Fe-Fe distance 2.8481 $\AA$ in
LaFeAsO$_{1-x}$F$_x$ \cite{sefa}. \\

{\bf B. Electrical resistivity analysis}\\

The measured temperature dependence of resistivity for
PrFeAsO$_{0.6}$F$_{0.12}$ (S1) and PrFeAsO$_{0.7}$F$_{0.12}$ (S2)
samples in the temperature range of 45-550 K is illustrated in Fig.
2(a). For both the samples, $\rho$  decreases initially at a very
slow rate and then at a faster rate with decreasing temperature
until the superconducting onset temperature ($T_{c}^{on}$$\sim$50 K)
is reached. Below $T_{c}^{on}$,  $\rho$ drops sharply and becomes
zero at around 47 K. The inset displays the enlarged view of the
onset of superconductivity for sample S1. For this sample, the
transition width ${\Delta}$${T_c}$ is 2.7 K, where ${\Delta}$${T_c}$
is defined as the width of a transition between 10$\%$ and 90$\%$ of
the normal-state resistivity. For S1 sample, we have measured
field-cooled (FC) and zero-field-cooled (ZFC) magnetic
susceptibilities at $H$=50 Oe. The values of shielding and Meissner
fractions determined from FC and ZFC are 65 and 23 $\%$,
respectively at 5 K. The resistivity ratio $\rho$(300 K)/$\rho$(54
K) of these samples is about 13. The resistivity ratio and the
transition width are respectively larger and narrower as compared to
that reported for other fluorine-doped and oxygen-deficient PrFeAsO
samples \cite{ren1,ren4}. The larger resistivity ratio and narrower
transition width are the indications of high sample quality. We have
also determined the approximate value of residual resistivity for
S1, $\rho(0)$$\sim$0.10 m$\Omega$ cm, by extrapolating the $\rho$ vs
$T$ curve between 150 and 60 K to $T$=0.  It may be mentioned that
we have prepared another sample with nominal oxygen content 0.50.
The value of resistivity ratio for this sample is also large
($\sim$9) and its $T_c$ is almost same as that for S1 and S2.
However, x-ray diffraction shows the presence of impurity phases in
this sample. Kito {\it et al}. \cite{hiro} prepared Nd-based
superconducting samples using high-pressure synthesis technique and
found that the amount of impurity phases increase rapidly when the
oxygen vacancy exceeds 0.4. This suggests that samples with oxygen
deficiency more than 0.4 can not be prepared in single phase. Though
both the magnitude and $T$ dependence of normal-state resistivity
depend to some extent on composition and purity of the samples,
$T_c$ is more or less insensitive to oxygen content. This
observation is consistent with the reported phase diagram $T_c$ vs
fluorine or oxygen content \cite{kami1,hiro}.\\

The analysis of $T$ dependence of $\rho$ reveals three different
temperature regimes. $\rho$ exhibits a quadratic temperature
dependence, $\rho$=$\rho(0)$+$AT^2$, in the range 70 K$\leq
T\leq$150 K while it is linear in $T$ in the intermediate region,
170 K$\leq T\leq$270 K. The $T^2$ behavior of $\rho$ below 150 K
indicates a strong electronic correlation and is consistent with the
formation of a Fermi-liquid state. Above 270 K, the resistivity
behavior shows marked departure from the usual temperature
dependence of simple metals and high-$T_c$ superconductors.
Normally, in good metals, $\rho$ is linear in $T$ at high
temperature due to the e-ph scattering.  For the present samples,
the slope  of $\rho$($T$) curve decreases continuously with
increasing temperature above 270 K and a saturationlike behavior
appears at higher temperature [Fig. 2(b)]. For sample S1, the
increase of $\rho$ is small while  $\rho$ is effectively
$T$-independent for sample S2 above 500 K [Fig. 2(c)]. The
saturationlike behavior of $\rho$ at high temperature at once draw
our attention to the $T$ dependence of $\rho$ for Nb, Chevrel
phases, several intermetallic and  A15 compounds \cite{nava}. This
phenomenon has been explained on the basis of the
conduction-electron mean free path approaching a lower limit with
the consequent breakdown of the classical Boltzmann theory
\cite{allen1}. The breakdown has been interpreted in terms of the
Ioffe-Regel criterion \cite{ioffe}. The electron mean free path can
not be shorter than the interatomic distance. The resistivity of
poor metals at high temperatures tends to saturate to a
temperature-independent value when the mean free path $l$ approaches
the wavelength $\lambda_{F}$=2$\pi/k_F$ associated with the Fermi
level, where $k_F$ is the Fermi wave vector. The Ioffe-Regel
criterion for the onset of this saturation is $k_F l$$\leq$1. In the
latter section, we will show
that this criterion is indeed satisfied in the present case.\\

Figure 3 shows the temperature dependence of resistivity for the
undoped PrFeAsO sample.  One can see that the behavior of
$\rho$($T$) for the parent sample is very different from that of its
superconducting counter part. An anomalous peak  at $T_s$$\sim$155 K
in $\rho$ is associated with the structural phase transition from
tetragonal to orthorhombic with decreasing temperature. Below $T_s$,
resistivity data can be fitted well with an expression,
$\rho$=$\rho(0) +aT^n$, with $n$ close to 1.5. However, $\rho$
increases linearly with $T$ from 250 K up to 525 K at the rate of
$\sim$ 1.5 $\mu$$\Omega$ cm/K. No saturationlike behavior is
observed for this sample in the measured temperature range. Thus,
the behavior of $\rho$ at high temperature for the
nonsuperconducting sample is quite different from that of the
superconducting sample. In the linear region of $\rho$ vs $T$
curves, the slope for the nonsuperconducting parent compound is
about 6 times smaller than that for the superconducting sample
($\sim$8.6 $\mu$$\Omega$ cm/K). \\

For the quantitative understanding of the behavior of $\rho(T)$ at
high temperature, we  estimate several normal-state transport
parameters related to the e-ph scattering and examine its effect on
superconductivity as it was done in the case of cuprate
superconductors \cite{gur}. Expressing  resistivity in terms of
plasma energy ($\hbar\omega_p$), $\rho(T)$=
4$\pi/(\omega_p^2\tau)=4\pi{v_F}/(\omega_p^2{l})$, where $v_F$ is
the Fermi velocity \cite{allen2}. At high temperature, where
resistivity is linear in $T$, $\tau$ is the e-ph scattering time
$\tau_{ep}$, and is given by $\hbar/\tau_{ep} =2\pi\lambda_{tr}{kT}$
\cite{allen2}. $\lambda_{tr}$ is closely related to the coupling
constant $\lambda$ that determines the superconducting transition
temperature. In the case of Nb and Pd it is found that
$\lambda_{tr}$ differs from $\lambda$ by about 10$\%$ \cite{nava}.
From the above relations one can deduce

\begin{equation}
\lambda_{tr}={\frac{\hbar\omega_p^2}{8\pi^2k}}{\frac{d\rho}{dT}}
\end{equation}

and

\begin{equation}
l={\frac{{\hbar}v_F}{2\pi \lambda_{tr} k T}}.
\end{equation}

We can calculate $\lambda_{tr}$  using the values of resistivity
slope (8.6 $\mu\Omega$ cm/K) in the linear region and plasma energy
($\sim$ 0.8 eV) determined from the inplane penetration depth
\cite{drew}. Eq. (1) reveals $\lambda_{tr}$ $=$ 1.3. Such a large
value of $\lambda_{tr}$ suggests that the e-ph coupling strength in
this system is quite strong. Using this value of $\lambda_{tr}$ and
Fermi velocity $v_F$=1.3$\times$$10^7$ cm/s calculated from band
theory \cite{sing}, we find $l$$\sim$ 2.5 $\AA$ at 500 K. This value
of $l$ is comparable to the Fe-Fe separation. We observe that the
Ioffe-Regel criterion for the onset of  saturation $k_F l$$\leq$1 is
satisfied in this case. $k_F$ was deduced using the values of Fermi
energy ($E_F$=0.4 eV) and Fermi velocity, as well as from the
experimental value of carrier density ($n$=10$^{21}$ cm$^{-3}$)
reported for LaFeAsO$_{1-x}$F$_{x}$ with free electron approximation
\cite{sing,sefa}. In the case of layered system, one can also check
the Ioffe-Regel criterion solely from the resistivity data using the
relation $k_F l$$\sim$($hc/4e$$^2$)/$\rho$, where $c$ is the
interlayer distance. If we use the value of $\rho$ at 500 K and
$c$=8.6 $\AA$, then also $k_F l$$<$1. Thus, the e-ph scattering is
dominating the high temperature resistivity and the large value of
$\lambda_{tr}$ is consistent with the resistivity saturation.\\

Having acquired  the qualitative and quantitative knowledge on
normal-state transport properties, we now discuss the role of e-ph
coupling on superconductivity. In the limit of strong e-ph coupling,
one can use the McMillan equation \cite{mcm} to estimate the
superconducting transition temperature

\begin{equation}
T_c={\frac{\Theta_D}{1.45}}
exp[-{\frac{1.04(1+\lambda)}{\lambda-\mu(1+0.62\lambda)}}].
\end{equation}

where $\mu$ is the Coulomb pseudopotential and $\Theta_D$ is the
Debye temperature. Assuming $\mu$$\sim$0 and
$\lambda$=$\lambda_{tr}$=1.3 and using the reported value of
$\Theta_D$=355 K for PrFeAsO \cite{kimber} we find that $T_c$=41 K.
This value of $T_c$ is only few Kelvin lower than the observed $T_c$
for PrFeAsO$_{0.6}$F$_{0.12}$. Nevertheless, the estimated and
observed values of $T_c$ are comparable in this class of materials.
A more accurate estimation of $T_c$ can be made using the
Allen-Dynes equation \cite{allen3} where the prefactor
${\Theta_D}/{1.45}$ in Eq. (3) is replaced by $\omega_{ln}$/1.2. As
we do not have  any knowledge on the value of $\omega_{ln}$ for
PrFeAsO, $T_c$ can not be deduced using this equation for the
present system. If we assume that e-ph interaction does not change
significantly from system to system then using the same value 1.3
for  $\lambda_{tr}$  along with theoretically derived
$\omega_{ln}$=206 K for LaFeAsO \cite{boer}, surprisingly the
observed value of $T_c$=27 K for LaFeAsO$_{0.89}$F$_{0.11}$ can be
reproduced. Now, we would like to comment on the dependence of $T_c$
on the rare earth ionic size in $R$FeAsO$_{1-x}$F$_{x}$. In
oxypnictide superconductors, it has been reported that the lattice
parameter $a$ decreases while $T_c$ increases when La is replaced by
smaller rare earth ions \cite{ren4}. The decrease of $a$ means the
decrease of Fe-Fe distance. Thus, one expects an increase in
characteristic phonon frequency and hence $T_c$ with the decrease of
ionic radius of $R$. This observation is consistent with the
phonon-mediated superconductivity but quite different from that of
cuprate superconductors where $T_c$ is not sensitive to the rare
earth ionic size. \\

It may be interesting to compare $\lambda_{tr}$ deduced from the
resistivity data with the theoretically predicted values for
understanding the role of e-ph interaction in Fe-based oxypnictides
and similar materials. In most of the reports, these Fe-based
materials are commonly viewed as unconventional superconductors
because of their high transition temperature in spite of the
predicted weak e-ph coupling and the occurrence of superconductivity
in close proximity to magnetism \cite{sing,lebe,haul,boer,mazi}. For
the undoped compound, the value (0.21) of the e-ph coupling strength
calculated from band theory \cite{boer} is comparable with that
(0.24) obtained from the resistivity data. However, there is a large
disagreement in the value of $\lambda_{tr}$ determined from the
resistivity data with the theoretical prediction for the
superconducting sample. According to the band structure calculation,
the value of $\lambda_{tr}$ does not change significantly as one
goes from the undoped to the doped sample, though, a 5-6 times
larger coupling constant ($\lambda_{tr}$=1.0-1.3) is needed to
reproduce the experimental value of $T_c$ in this system
\cite{boer}. On the contrary, if $\lambda_{tr}$ does not change
significantly, it is not clear why the behavior of resistivity
changes so dramatically with a small amount of fluorine doping.
Eschrig  \cite{esch} pointed out several shortcomings of the
standard band structure calculations and argued that the e-ph
coupling in this system is quite strong due to the Fe inplane
breathing mode. Drechsler {\it et al}. \cite{drec} also cast doubt
on the calculated weak value of e-ph coupling. The gap function is
the single most important quantity that can be used to reveal the
pairing mechanism of a superconductor. Chen {\it et al}. \cite{chen}
studied Andreev spectroscopy on SmFeAsO$_{0.85}$F$_{0.15}$ and
observed a single nodeless gap with nearly isotropic in size across
different sections of the Fermi surface. The value of the normalized
gap parameter (2$\Delta/kT_c$) determined by them was 3.68 which is
slightly larger than the BCS prediction of 3.52 in the weak coupling
regime but close to that observed in many conventional
superconductors with strong e-ph coupling such as Nb, V$_3$Si, etc
\cite{chen,pool}. Also, the $T$ dependence of the gap is consistent
with the BCS prediction, but dramatically different from that of the
pseudogap in the cuprate superconductors. From these observations,
they concluded that the structure of gap is not compatible with
theoretical models involving antiferromagnetic fluctuations, strong
electronic correlations, the $t$-$J$ model, and that proposed for
the cuprate superconductors. Isotropic $s$-wave symmetry of the gap
has also been established from the angle resolved photo emission
spectroscopy \cite{kondo}. It has been argued that the isotropic
nature of gap and the appearance of superconductivity in close
proximity to a suppressed structural phase transition together with
the observation of superconductivity in CaFe$_2$As$_2$ with
application of a modest pressure bring the role of phonon in these
Fe-based samples to the forefront \cite{kondo}.\\

The strong deviation of resistivity from linearity to saturationlike
in superconducting oxypnictides immediately draw our attention to
conventional superconductors  where e-ph coupling is strong such as
in A15 compounds and Chevrel phases \cite{nava}. In these cases,
$\rho$ shows a saturationlike behavior at high temperatures and the
values of $\lambda_{tr}$, in general, are quite large and comparable
with that of the present system \cite{pool}. It may be important to
mention that apart from $\lambda_{tr}$, other parameters related to
superconductivity such as the density of states at the Fermi level,
the magnitude and $T$ dependence of 2$\Delta/kT_c$ and the symmetry
of gap-parameter of these systems are comparable to oxypnictides
\cite{pool}. The $T$ dependence of gap parameter is compatible with
the BCS prediction in both the cases. So by comparing these systems
with the present one it is apparent that oxypnictides  belong
to a class of strongly coupled BCS superconductors.\\

{\bf IV. CONCLUSIONS}

In conclusion, we have analyzed the temperature dependence of
resistivity of PrFeAsO$_{1-x}$F$_y$ samples. For nonsuperconducting
PrFeAsO, $\rho$ below 155 K shows a power-law behavior whereas in
the high temperature region (250-525K), $\rho$ is linear in $T$. For
superconducting samples, $\rho$ above $T_c$ crosses over from $T^2$
dependence due to the electron-electron interaction to linear in $T$
and then to a saturationlike behavior at higher temperature due to
the e-ph interaction. The resistivity saturation indicates that the
e-ph interaction is strong and the conduction-electron mean free
path approaching a lower limit with the consequent breakdown of the
classical Boltzmann theory. We have estimated the e-ph coupling
parameter $\lambda_{tr}$ from the linear dependence of $\rho$ to be
approximately 0.24 and 1.3, respectively for nonsuperconducting and
superconducting samples. The small value of $\lambda_{tr}$ for the
former is consistent with band structure calculations, while for the
latter, it is 5-6 times larger than the theoretical value. The
present resistivity results along with the structure, value and
temperature dependence of the gap-parameter 2$\Delta/kT_c$, and
other important parameters related to superconductivity suggest that
Fe-based oxypnictides are a class of BCS superconductors with strong
e-ph coupling similar to
Chevrel phases and A15 compounds.\\

{\bf ACKNOWLEDGMENTS}

The authors would like to thank B. Ghosh and A. N. Das stimulating
and A. Pal for the technical help during the sample preparation and
measurements.\\

\newpage

\begin{figure}[htb]
\begin{center}
\includegraphics[scale=2.3,angle=0]{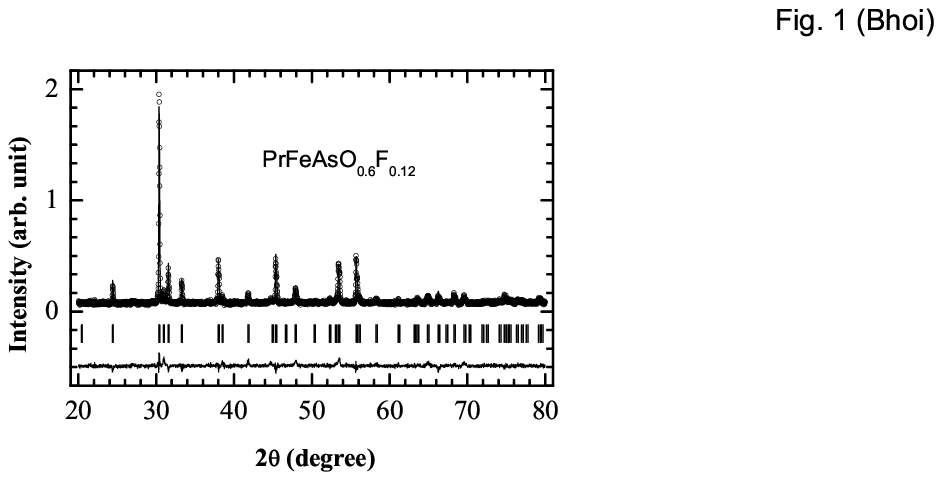}
\caption{ The x-ray diffraction pattern for the superconducting
PrFeAsO$_{0.6}$F$_{0.12}$  sample. The solid line corresponds to
Rietveld refinement of the diffraction pattern with $P4/nmm$ space
group.} \label{Fig1}
\end{center}
\end{figure}

\begin{figure}[htb]
\begin{center}
\includegraphics[scale=1.3,angle=0]{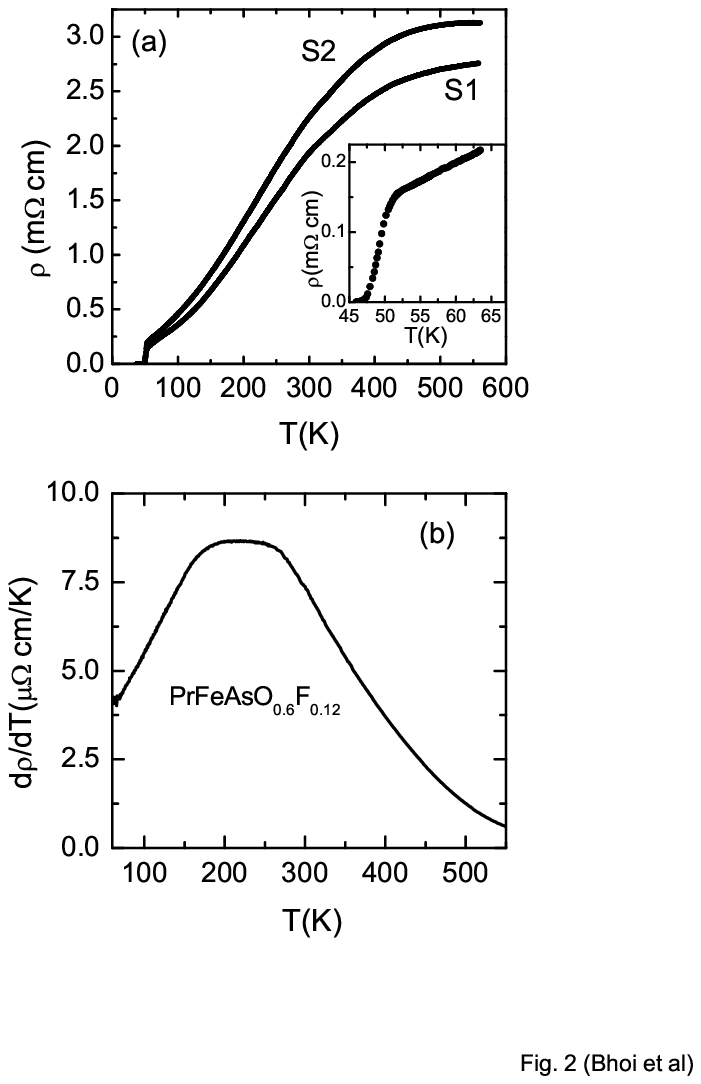}
\caption{ Temperature dependence of (a) resistivity ($\rho$) for the
superconducting PrFeAsO$_{0.6}$F$_{0.12}$ (S1) and
PrFeAsO$_{0.7}$F$_{0.12}$ (S2) samples, and (b) $d\rho/dT$ for
PrFeAsO$_{0.6}$F$_{0.12}$ sample. Inset of (a) is the enlarged view
of resistivity change close to the superconducting transition
temperature for PrFeAsO$_{0.6}$F$_{0.12}$.} \label{Fig2}
\end{center}
\end{figure}

\begin{figure}[htb]
\begin{center}
\includegraphics[scale=2.3,angle=0]{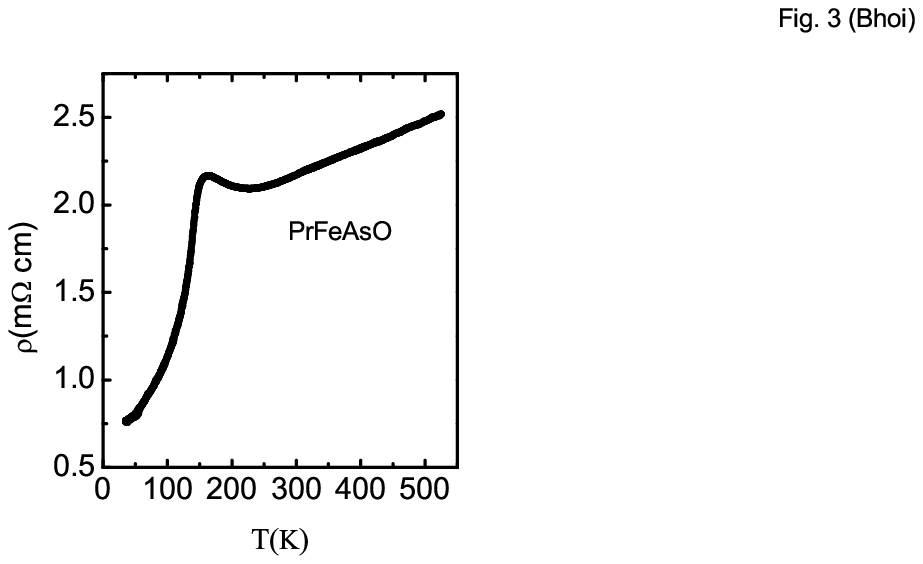}
\caption{Temperature dependence of resistivity  for the
nonsuperconducting PrFeAsO sample.} \label{Fig3}
\end{center}
\end{figure}

\begin{references}

\bibitem{kami1} Y. Kamihara, T. Watanabe, M. Hirano,  and H. Hosono,  J. Am.
Chem. Soc. {\bf 130}, 3296 (2008).

\bibitem{kami2} H. Takahashi, K. Igawa, K. Arii, Y.
Kamihara, M. Hirano, and H. Hosono,   Nature (London) {\bf 453}, 376
(2008).

\bibitem{chen1} X. H. Chen, T. Wu, G. Wu, R. H. Liu, H. Chen, and D. F.
Fang,  Nature (London) {\bf 453}, 761 (2008).

\bibitem{chen2} G. F. Chen, Z. Li, D. Wu, G. Li, W. Z. Hu, J.
Dong, P. Zheng, J. L. Luo, and N. L. Wang, Phys. Rev. Lett. {\bf
100}, 247002 (2008).

\bibitem{ren1} Z. A. Ren, J. Yang, W. Lu, W. Yi, G. C. Che, X. L. Dong,
L. L. Sun, and Z. X. Zhao, Materials Research Innovations, {\bf
12},1,(2008).

\bibitem{ren2} Z. A. Ren, J. Yang, W. Lu, W. Yi, X. L. Shen, Z. C. Li,
G. C. Che, X. L. Dong, L. L. Sun, F. Zhou, and Z. X. Zhao, Euro.
Phys. Lett. {\bf 82}, 57002 (2008).

\bibitem{ren3} Z. A. Ren, W. Lu, J. Yang, W. Yi, X. L. Shen, Z. C. Li,
G. C. Che, X. L. Dong, L. L. Sun, F. Zhou, and Z. X. Zhao, Chin.
Phys. Lett. {\bf 25}, 2215 (2008).

\bibitem{dong} J. Dong et al., Euro. Phys. Lett. {\bf 83}, 27006 (2008).

\bibitem{chen} T. Y. Chen, Z. Tesanovic, R. H. Liu, X. H. Chen, and
C. L. Chien, Nature (London) {\bf 453}, 1224 (2008).

\bibitem{kondo} T. Kondo, et al. arxiv:0807.0815v1.

\bibitem{drec} S.-L. Drechsler et al. arxiv:0805.1321v1.

\bibitem{sing} D. J. Singh and M. H. Du, Phys. Rev. Lett. {\bf 100}, 237003
(2008).

\bibitem{lebe} S. Lebegue, Phys. Rev. B {\bf 75}, 035110 (2007).

\bibitem{haul} K. Haule, J. H. Shim, and G. Kotliar, Phys. Rev. Lett. {\bf 100},
226402 (2008).

\bibitem{boer} L. Boeri, O. V. Dolgov, and A. A. Golubov, Phys. Rev. Lett. {\bf 101},
026403 (2008).

\bibitem{mazi} I. I. Mazin, Phys. Rev. B {\bf 75}, 094407 (2007).

\bibitem{esch} H. Eschrig, arxiv:0804.0186v2.

\bibitem{ren4} Z. A. Ren, G. C. Che, X. L. Dong, J. Yang, W. Lu, W. Yi,
X. L. Shen, Z. C. Li, L. L. Sun, F. Zhou, and Z. X. Zhao, Euro.
Phys. Lett. {\bf 83}, 17002 (2008).

\bibitem{hiro} H. Kito, H. Eisaki, and A. Iyo, J. Phys. Soc. Jpn. {\bf 77},
063707 (2008).

\bibitem{sefa} A. S. Sefat, M. A. McGuire, B. C. Sales,
R. Jin, J. Y. Howe, and D. Mandrus, Phys. Rev.  B {\bf 77}, 174503
(2008).

\bibitem{ding} L. Ding, C. He, J. K. Dong, T. Wu, R. H. Liu, X. H.
Chen, and S. Y. Li, Phys. Rev. B {\bf 77}, 180510(R), (2008).

\bibitem{nava} F. Nava, O. Bsi and K. N. Tu Phys. Rev. B {\bf 34}, 6134, (1986)
and references therein.

\bibitem{allen1} P. B. Allen and B. Chakraborty, Phys. Rev. B {\bf 23},
4815 (1981).

\bibitem{ioffe} A. F. Ioffe and A. R. Regel, Prog. Semicond. {\bf 4},
237 (1960).

\bibitem{gur} M. Gurvitch and A.T. Fiory   Phys. Rev. Lett.  {\bf 59}, 1337
(1987).

\bibitem{allen2} P. B. Allen, W. E. Pickett, and H. Krakauer, Phys. Rev.
B {\bf 37}, 7482 (1988).

\bibitem{drew} A. J. Drew, F. L. Pratt, T. Lancaster, S. J. Blundell,
P. J. Baker, R. H. Liu, G. Wu, X. H. Chen, I. Watanabe, V. K. Malik,
A. Dubroka, K. W. Kim, M. R\"{o}ssle, and C. Bernhard,
arXiv:0805.1042 (unpublished); R. Khasanov,  H. Luetkens, A. Amato,
H. H. Klauss, Z. A. Ren, J. Yang, W. Lu, and Z. X. Zhao,
arXiv:0805.1923 (unpublished).

\bibitem{mcm} McMillan, Phys. Rev. {\bf 167}, 331 (1968).

\bibitem{kimber} S. A. J. Kimber, D. N. Argyriou, F. Habicht, S.
Gevischer, T. Hansen, T. Chatterji, R. Klinger, C. Hess, G. Behr, A.
Kondrat, B. Buechner   arXiv:0807.4441 (unpublished)

\bibitem{allen3} P. B. Allen and R. C. Dynes, Phys. Rev. B {\bf 12}, 905 (1975).

\bibitem{pool} C. P. Poole, Jr. , H. A. Farach, R. J. Creswick,
   Superconductivity   (Academic Press, Inc.  San Diego, 1995 )
\end{references}
\end{document}